\documentclass[letterpaper]{article} 

\usepackage[preprint]{aaai2027}  
\usepackage[hyphens]{url}  
\usepackage{graphicx} 
\usepackage{natbib}  
\usepackage{caption} 
\usepackage{algorithm}
\usepackage{algorithmic}

\usepackage{newfloat}
\usepackage{amsmath}
\usepackage{listings}
\usepackage{xspace}
\newcommand{\Name}{$\mathtt{InkShield}$\xspace}

\DeclareCaptionStyle{ruled}{labelfont=normalfont,labelsep=colon,strut=off} 
\floatstyle{ruled}
\newfloat{listing}{tb}{lst}{}
\floatname{listing}{Listing}

\usepackage{booktabs}

\title{\Name: Writing Style Protection Against Unauthorized Handwriting Mimicry}

\affiliations{
    University of Electronic Science and Technology of China\\
    j.xiong@std.uestc.edu.cn,
    wenbo\_jiang@uestc.edu.cn
}

\author{
  Jian Xiong\textsuperscript{1},
  Wenbo Jiang\textsuperscript{1}\corresponding,
  Zihan Wang\textsuperscript{1},
  Rui Zhang\textsuperscript{1},\\
  Wenshu Fan\textsuperscript{1},
  Hongwei Li\textsuperscript{1},
  Guowen Xu\textsuperscript{1}
}
\affiliations{
    \textsuperscript{\rm 1} University of Electronic Science and Technology of China \\
    j.xiong@std.uestc.edu.cn,
    wenbo\_jiang@uestc.edu.cn
}

\begin{document}

\maketitle

\begin{abstract}

Recent handwritten text generators can reproduce a writer's style from publicly available references, posing risks of document forgery and identity misuse. An attacker may use a publicly available handwritten note or signature sample to generate forged recommendation letters or authorization forms, leading to document fraud, identity misuse, and misleading decisions. 
However, existing protections against unauthorized image editing or synthesis transfer poorly to handwriting style mimicry. Designed for natural images with complex backgrounds, they often optimize perturbations over the whole image. For sparse handwriting images, such global perturbations become conspicuous in blank background regions and largely degrade the visual quality.
In this work, we propose \Name, a proactive writing-style defense that protects reference images before release. \Name selects a decoy writer to define a style-displacement direction, optimizes perturbations with a frozen handwriting-generation surrogate, and confines them to ink-stroke edges to avoid conspicuous background artifacts. On IAM, the average Top-1/Top-5 rates at which generated samples are retrieved as the target writer by two independent writer evaluators decrease from $11.94\%$/$36.52\%$ to $2.03\%$/$8.79\%$. Meanwhile, the protected references remain visually close to the originals (LPIPS $0.0078$), and the generated text remains readable. \Name also exhibits transferability to other handwriting generators. Overall, \Name provides practical protection against unauthorized handwriting style mimicry.

\end{abstract}

\section{Introduction}

Handwritten text generation models can synthesize new words or phrases with controllable textual content and writer style~\cite{kang2020ganwriting,bhunia2021handwriting,pippi2023visual,nikolaidou2023wordstylist}. This capability is useful for data augmentation, font-like personalization, and historical document restoration. Moreover, recent SOTA generators can achieve one-shot mimicry, reproducing a target writer's style from only a single reference sample~\cite{dai2024onedm,le2026constant}, thereby significantly lowered the barrier. However, this strong capability also introduces a new security risk: a single handwriting sample can become a reusable style key. A scanned form, a homework sheet, or a handwritten note, once used as a reference image, may facilitate document forgery, identity misuse, and other serious harms.

\begin{figure}[!t]
    \centering
    \includegraphics[width=0.98\columnwidth]{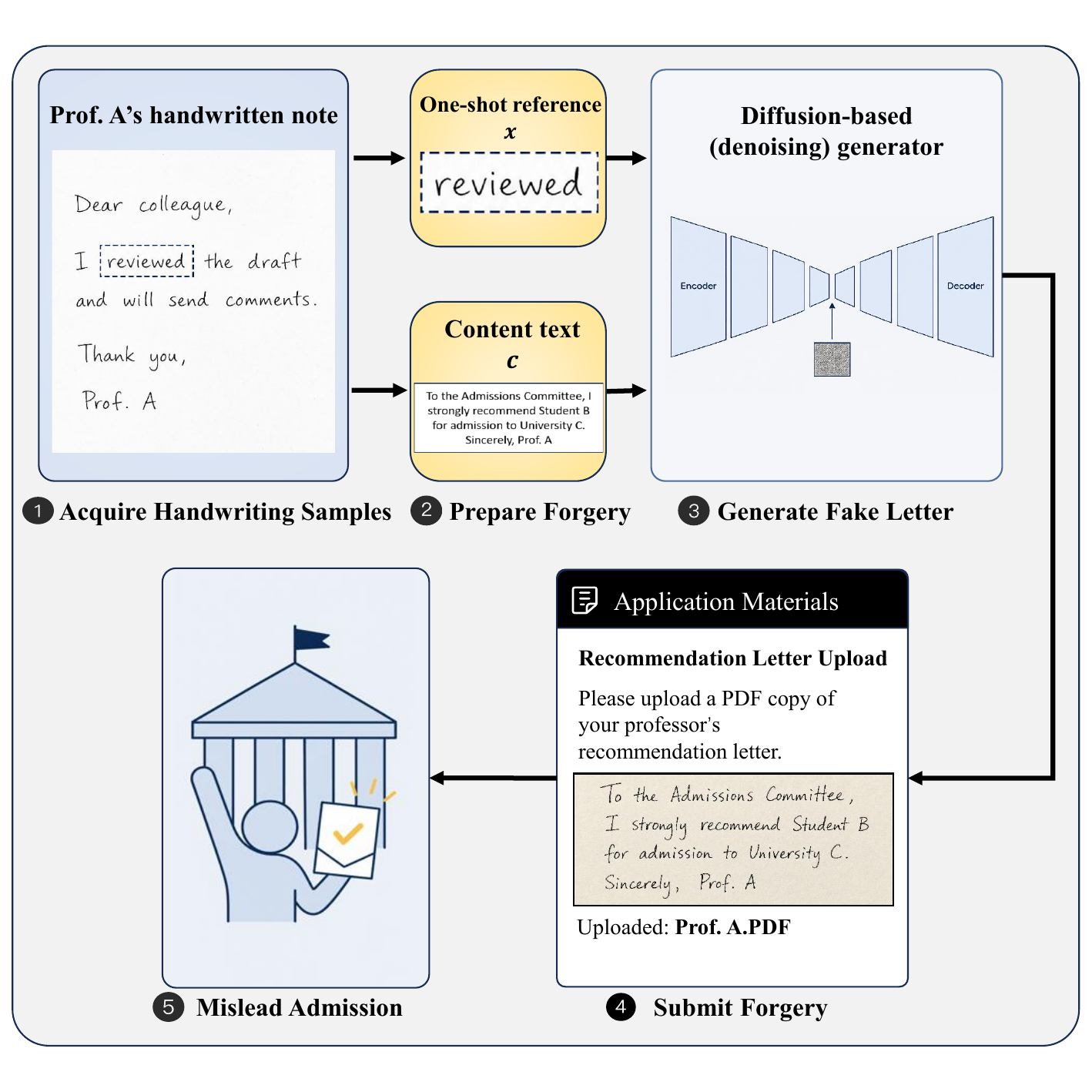}
    \caption{\textbf{Security risk of one-shot handwriting style mimicry:} A public handwritten word enables an attacker to generate and submit a forged recommendation letter in the professor's style.}
    \label{fig:privacy_risk}
\end{figure}

Figure~\ref{fig:privacy_risk} illustrates the threat posed by one-shot handwriting style reuse. With only a single handwriting reference image, the attacker can synthesize new content in a style similar to that of the target writer. However, for practical purposes such as submitting coursework, sharing notes, or showcasing calligraphic works, reference owners may inevitably need to upload handwritten materials online. This raises a reference-side protection problem: \noindent\textit{Can we protect handwriting images before release while preserving their readability and weakening downstream target-writer mimicry?}

Existing image-protection methods primarily target natural-image scenarios, including unauthorized image editing, face or artistic-style imitation~\cite{salman2023photoguard,choi2025diffusionguard,vanle2023antidreambooth,liu2024metacloak,li2025styleguard}. When directly applied to handwriting references, these methods face two limitations. First, their objectives are not optimized for handwriting style protection. Because writer-specific style is expressed through fine-grained local stroke morphology, generic image-level perturbation objectives may provide limited suppression of target-writer mimicry. Second, they often optimize perturbations globally over the entire image, introducing noticeable artifacts in the largely blank background regions of handwritten word images and degrading visual quality. 

To address these two challenges, we propose \Name, a reference-side writing-style protection method against unauthorized handwriting mimicry. Before release, \Name proactively applies low-visibility perturbations to protect handwriting references, impairing downstream mimicry while preserving readability and visual utility. Specifically, \Name consists of three stages: decoy writer selection, handwriting-aware perturbation constraints, and Surrogate-Guided Generative Disruption. A decoy pool provides a style-displacement direction that pushes the protected reference away from the target-writer style, addressing the limited ability of generic image-level objectives to disrupt writer-specific style. A handwriting-aware stroke-edge mask confines perturbations to style-sensitive regions, avoiding noticeable artifacts in blank background areas caused by global perturbations. Finally, a frozen surrogate generation objective uses multiple pseudo-content words during optimization, allowing the learned perturbation to generalize across different target texts rather than overfitting to a particular word.


In our experiments, we evaluate \Name on the IAM dataset from four complementary perspectives: target-writer mimicry suppression, reference stealth, content preservation, and generation quality. Under the primary One-DM (one-shot) setting, \Name reduces the average target-writer Top-5 retrieval rate across two independent writer evaluators from $36.52\%$ to $8.79\%$, while achieving strong reference stealth with an LPIPS of $0.0078$. Results on CONSTANT(one-shot) and DiffusionPen(multi-shot) further demonstrate cross-model transferability and effectiveness with multiple handwriting references.

Our contributions are summarized as follows:
\begin{itemize}
\item We formulate unauthorized generative handwriting style mimicry as a reference-side protection problem and propose a proactive defense method for protecting released handwriting references. To the best of our knowledge, this is the first work to address this problem.

\item We propose \Name, which integrates decoy-guided style displacement, handwriting-aware stroke-edge constraints, and frozen surrogate-guided generative disruption. It suppresses target-writer style reuse while preserving the readability and visual utility of released handwriting references.


\item We conduct extensive experiments on One-DM, CONSTANT, and DiffusionPen. Results show effective mimicry suppression with preserved reference readability, and transferability across generators.

\end{itemize}

\begin{figure*}[!t]
    \centering
    \includegraphics[width=\textwidth]{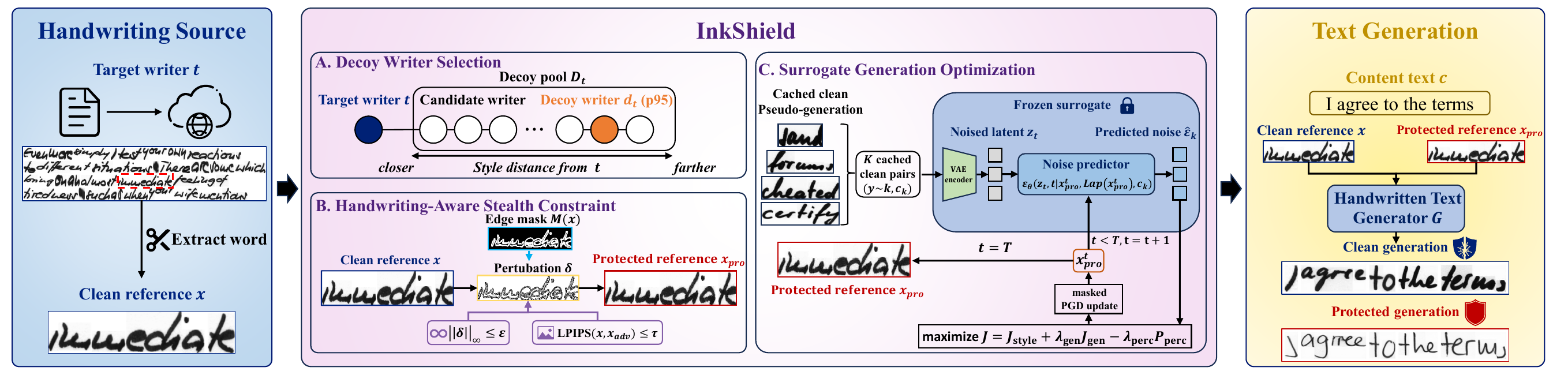}
    \caption{\textbf{Overview of \Name.} Given a clean reference from a target writer, \Name selects a distant decoy writer to define a style-displacement direction, confines low-visibility perturbations to stroke-edge regions, and optimizes them through a frozen surrogate generation chain over multiple pseudo-content words. The protected reference reduces downstream target-writer mimicry while preserving content readability and visual quality.}
    \label{fig:sap_overview}
\end{figure*}

\section{Related Work}

\subsection{Handwritten Text Generation}

Handwritten text generation synthesizes text images with controllable content and writer style. A central distinction among existing methods is whether they operate in multi-shot or one-shot settings. Multi-shot methods require collecting multiple handwriting samples from a writer to obtain a more complete style representation. GANwriting performs content- and style-conditioned handwritten word generation in a multi-shot setting~\cite{kang2020ganwriting}. HWT uses Transformer attention to model global and local patterns from multiple style examples~\cite{bhunia2021handwriting}. VATr employs visual archetypes for multi-shot generation with rare characters and unseen writers~\cite{pippi2023visual}, while VATr++ improves input preparation, regularization, and evaluation for this setting~\cite{vanherle2024vatrpp}. DiffusionPen further formulates handwritten text generation as a five-shot latent-diffusion problem~\cite{nikolaidou2024diffusionpen}. In contrast, one-shot methods aim to reproduce a writer's style from only one handwriting image. HiGAN performs one-shot handwriting imitation through content--style disentanglement~\cite{gan2021higan}. One-DM enhances style extraction from a single reference using high-frequency information~\cite{dai2024onedm}, and CONSTANT improves one-shot generation through patch contrastive enhancement and style-aware quantization~\cite{le2026constant}. DiffBrush uses one-shot style learning, extending generation from isolated words to complete handwritten text lines by modeling intra- and inter-word style patterns~\cite{dai2025diffbrush}. 

\Name protects released handwriting references against unauthorized target-writer mimicry and remains effective in both one-shot and multi-shot settings, while primarily focusing on the stringent setting in which a single exposed reference is sufficient for mimicry.

\subsection{Protection Against Unauthorized Image Usage}

Existing work has explored protections against unauthorized uses of images, including style mimicry, subject personalization, and image editing. Glaze protects artworks from text-to-image style mimicry~\cite{shan2023glaze}. Anti-DreamBooth protects personal images from unauthorized diffusion personalization~\cite{vanle2023antidreambooth}, while MetaCloak improves the transferability of such protections through meta-learning~\cite{liu2024metacloak}. PAP considers both face privacy and artistic style in personalized diffusion models~\cite{wan2024pap}, and StyleGuard targets text-to-image style mimicry using style perturbations~\cite{li2025styleguard}. PhotoGuard, EditShield, and DiffusionGuard further protect images against diffusion-based editing~\cite{salman2023photoguard,chen2024editshield,choi2025diffusionguard}. Mist improves adversarial perturbation optimization for diffusion models~\cite{liang2023mist}. Despite their different downstream targets, these methods are primarily designed for natural images, artistic styles, or faces. When transferred directly to handwriting references, their generic image-level objectives may not sufficiently disrupt the fine-grained stroke morphology that encodes writer-specific style, while global perturbations can introduce noticeable artifacts in the largely blank backgrounds of handwritten word images. 

To address these limitations, \Name uses decoy-guided style displacement to interfere with writer-specific style extraction and confines perturbations to stroke-edge regions to preserve visual quality.

\section{Threat Model}

\noindent\textbf{Scenario.}
We prioritize a scenario in which a handwriting reference image is used without authorization for style mimicry in the one-shot setting. A target writer $t$ owns a clean handwriting reference image $x$ and needs to publish or share it online. After obtaining the released image, an attacker provides it together with arbitrary content text $c$ to a handwriting generator $G$ and synthesizes new content $G(c,x)$ in the style of the target writer.

\noindent\textbf{Attacker's Capability.}
The attacker can obtain a released handwriting reference image and use a
handwriting generator for style mimicry. The attacker may choose arbitrary
content text and, in the one-shot setting, requires only a single publicly
available reference image (e.g., a word image) to carry out the attack.

\noindent\textbf{Defender's Capability.}
Before releasing the reference image, the defender can access the clean image $x$ and apply low-visibility perturbations to obtain a protected image $x_{\mathrm{pro}}$ for publication. To construct this perturbation, the defender may use a surrogate handwriting generator and writer-style representations during optimization. However, the defender does not know the downstream generator, its parameters, or its processing pipeline used by the attacker, and cannot control the attacker's content text, random seed, or additional reference samples.

\noindent\textbf{Defender's Goals.}
The primary goal is target-writer mimicry suppression: when the attacker uses $x_{\mathrm{pro}}$ for generation, the resulting handwriting should no longer closely resemble the target writer's style. At the same time, the protected reference should preserve its original textual content and readability under low-visibility perturbations. A successful defense must not rely on destroying the released reference.

\section{Methodology}

We propose \Name to protect handwriting references before release. As shown in Figure~\ref{fig:sap_overview}, \Name selects a stylistically distant decoy writer from the target writer's decoy pool to define a style-displacement direction and optimizes low-visibility perturbations confined to stroke-edge regions through a frozen surrogate. In our implementation, this surrogate is instantiated with One-DM~\cite{dai2024onedm}. Algorithm~\ref{alg:inkshield} summarizes the complete protection workflow.

\begin{algorithm}[t]
\caption{\textbf{Algorithm for \Name}}
\label{alg:inkshield}
\small
\begin{algorithmic}[1]
\REQUIRE Clean reference $x$ of target writer $t$; authorized candidate
writers $\mathcal{W}$ and their reference samples; frozen surrogate
$G_{\theta}$; pseudo texts $C$; $\rho$, $\epsilon$, $\alpha$, and $T$
\ENSURE Protected reference $x_{\mathrm{pro}}$

\STATE Compute writer prototypes $\mathbf{p}_{t}$ and
$\{\mathbf{p}_{u}\}_{u\in\mathcal{W}}$ from available reference samples
\STATE Compute $\Delta(t,u)=1-\cos(\mathbf{p}_{t},\mathbf{p}_{u})$
for all $u\in\mathcal{W}$
\STATE Sort candidates by increasing $\Delta(t,u)$ to obtain the ordered
decoy pool $D_t$
\STATE Select $d_t\gets D_t[\lceil\rho|D_t|\rceil]$ and a fixed
reference $x_{d_t}$ of $d_t$
\STATE Construct the handwriting-aware edge mask $M(x)$ using
Eq.~\ref{eq:edge_mask}
\STATE Cache clean pseudo generations
$\{\widetilde{y}_{c}\gets G_{\theta}(c,x)\}_{c\in C}$
\STATE Initialize
$\delta^{(0)}\sim\mathcal{U}(-\epsilon,\epsilon)\odot M(x)$

\FOR{$k\gets 0$ \TO $T-1$}
    \STATE $x_{\mathrm{pro}}\gets
    \operatorname{clip}_{[0,1]}
    \left(x+M(x)\odot\delta^{(k)}\right)$
    \STATE Compute $J_{\mathrm{style}}$ using
    Eq.~\ref{eq:style_objective}
    \STATE Compute $J_{\mathrm{gen}}$ using
    Eq.~\ref{eq:generation_objective}
    \STATE Compute $P_{\mathrm{perc}}$ using
    Eq.~\ref{eq:perceptual_penalty}
    \STATE Compute the total objective $J$ using
    Eq.~\ref{eq:sap_objective}
    \STATE Update $\delta^{(k+1)}$ using Eq.~\ref{eq:pgd_update}
\ENDFOR

\RETURN $x_{\mathrm{pro}}\gets
\operatorname{clip}_{[0,1]}
\left(x+M(x)\odot\delta^{(T)}\right)$
\end{algorithmic}
\end{algorithm}

\subsection{Decoy Writer Selection}
\Name selects a decoy writer to guide the displacement of the protected reference in writer-style space. The selected decoy serves only as a defense-side displacement direction: the goal is to reduce similarity to the target writer rather than make downstream generations imitate the decoy writer. For each target writer $t$, the defender constructs a decoy pool $D_t$ comprising authorized candidate writers other than $t$ and their reference samples.

To measure the style distance between the target writer and the candidate writers in $D_t$, we draw on the style representation mechanism of One-DM. One-DM decomposes handwriting style into complementary low-level and high-level components, corresponding to fine-grained stroke appearance and more global writer-level characteristics, respectively. Following this observation, we extract low-level and high-level style representations from handwriting reference images, denoted by $\phi^{\mathrm{low}}(x)$ and $\phi^{\mathrm{high}}(x)$. For the target writer and each candidate writer $w$, we average these features over the writer's available handwriting samples to obtain component-wise prototypes $\mathbf{p}^{\mathrm{low}}_w$ and $\mathbf{p}^{\mathrm{high}}_w$, and concatenate them as the writer-level prototype $\mathbf{p}_w$. We then compute the style distance between the target writer $t$ and each candidate writer $u$:
\begin{equation}
\label{eq:style_distance}
\Delta(t,u)=1-\cos(\mathbf{p}_t,\mathbf{p}_u).
\end{equation}
We sort the candidate writers in $D_t$ by increasing $\Delta(t,u)$; after sorting, $D_t[j]$ denotes the $j$-th closest candidate writer to $t$, where $j\in\{1,\ldots,|D_t|\}$ and $|D_t|$ denotes the number of candidate writers in the pool.

Decoy selection balances perturbation cost against mimicry suppression. A nearby decoy may require a smaller perturbation and thus better preserve the visual quality of the protected reference, but its style remains close to that of the target writer and may provide limited suppression. In contrast, a distant decoy provides a larger style-displacement direction and can more effectively reduce the similarity between generated handwriting and the target-writer style. However, an excessively distant decoy may introduce unstable guidance or require stronger perturbations. Our decoy-pool sensitivity study shows that distant decoys generally yield stronger target-writer mimicry suppression. We therefore select a high-percentile decoy writer $d_t$ from the ordered pool:
\begin{equation}
\label{eq:decoy_selection}
d_t =
D_t\!\left[
\left\lceil
\rho |D_t|
\right\rceil
\right],
\end{equation}
where $\rho$ is the decoy percentile. We use $\rho=0.95$ as the default setting, which achieves strong mimicry suppression without relying on the potentially outlying farthest writer.

\subsection{Handwriting-Aware Stealth Constraint}

Handwritten word images are sparse: foreground strokes occupy only a small portion of the image, whereas most pixels belong to the blank background. Meanwhile, writer-specific style is encoded in local stroke morphology, including stroke width, curvature, slant, junctions, and boundary shapes. \Name therefore constrains perturbations to a handwriting-aware mask that focuses on stroke-boundary regions.

Given a grayscale reference image $x\in[0,1]$, we first obtain an ink foreground mask $F(x)=\mathbf{1}[x<\eta_{\mathrm{fg}}]$, where $\eta_{\mathrm{fg}}=0.92$. Let $\mathcal{D}_{1}(\cdot)$ and $\mathcal{E}_{1}(\cdot)$ denote morphological dilation and erosion with a one-pixel radius, respectively. We construct a morphological stroke boundary as
\begin{equation}
B(x)
=
\left[
\mathcal{D}_{1}(F(x))
-
\mathcal{E}_{1}(F(x))
\right]_{+},
\end{equation}
where $[\cdot]_{+}$ clips negative values to zero. In addition, we compute the normalized gradient magnitude $S_{\mathrm{sob}}(x)$ using a $3\times3$ Sobel operator~\cite{sobel1968isotropic} and retain strong responses as $H(x)=\mathbf{1}[S_{\mathrm{sob}}(x)>\eta_{\mathrm{sob}}]$, where $\eta_{\mathrm{sob}}=0.12$. We retain only the strong Sobel responses within a one-pixel neighborhood of the ink foreground and take their binary union with $B(x)$ to form a preliminary edge support $E(x)$. The final mask is then defined as
\begin{equation}
\label{eq:edge_mask}
M(x) = F(x)\odot\mathcal{D}_{1}\big(E(x)\big).
\end{equation}
This construction expands the combined edge support by one pixel and finally restricts it to the ink foreground, concentrating perturbations near stroke boundaries while avoiding the blank background.

Given the clean reference image $x$, the protected reference is defined as
\begin{equation}
\label{eq:protected_reference}
x_{\mathrm{pro}}=\operatorname{clip}_{[0,1]}\left(x + M(x)\odot\delta\right),\qquad \|\delta\|_{\infty}\le\epsilon,
\end{equation}
where $\delta$ is the perturbation, $\odot$ denotes element-wise multiplication, and $\epsilon$ controls the perturbation budget.

In addition to the spatial mask and the $\ell_{\infty}$ constraint, \Name uses a soft perceptual penalty to encourage stealthiness. We measure perceptual distortion using LPIPS~\cite{zhang2018unreasonable}:
\begin{equation}
\label{eq:perceptual_penalty}
P_{\mathrm{perc}}=\max\left(0,\operatorname{LPIPS}(x_{\mathrm{pro}},x)-\tau\right),
\end{equation}
where $\tau$ is a penalty threshold rather than a strict upper bound. This term discourages perceptually noticeable changes while allowing the perturbation to suppress target-writer mimicry.

During optimization, we mask the gradient updates and project $\delta$ onto the $\ell_{\infty}$ ball after each step. The multiplication by $M(x)$ in Eq.~\ref{eq:protected_reference} further guarantees that the final perturbation is confined to the handwriting-aware region. 

\subsection{Surrogate-Guided Generative Disruption}

Having defined the decoy-guided displacement direction and handwriting-aware perturbation constraints, \Name optimizes the perturbation $\delta$ through a surrogate handwriting generator to suppress target-writer mimicry in downstream generations.

We first define a style displacement objective using the selected decoy writer. During optimization, \Name uses the clean reference $x$ of the target writer $t$ and a fixed representative reference $x_{d_t}$ from the selected decoy writer $d_t$. 
Let $r\in\{\mathrm{low},\mathrm{high}\}$ denote the low-level and high-level style components. We define
\begin{equation}
\label{eq:style_objective}
\begin{aligned}
J_{\mathrm{style}}
=
\sum_{r\in\{\mathrm{low},\mathrm{high}\}}
\Big[
&\lambda_{\mathrm{dec}}^{r}
\cos\!\left(
\phi^{r}(x_{\mathrm{pro}}),
\phi^{r}(x_{d_t})
\right) \\
&-
\lambda_{\mathrm{tar}}^{r}
\cos\!\left(
\phi^{r}(x_{\mathrm{pro}}),
\phi^{r}(x)
\right)
\Big].
\end{aligned}
\end{equation}
where $\lambda_{\mathrm{dec}}^{r}$ and $\lambda_{\mathrm{tar}}^{r}$ control decoy attraction and target-writer repulsion at style level $r$, respectively. Thus, the objective moves the protected reference toward a decoy-guided direction while explicitly moving it away from the target-writer style.

Inspired by Fully-trained Surrogate Model Guidance (FSMG) in Anti-DreamBooth~\cite{vanle2023antidreambooth}, we use a frozen One-DM~\cite{dai2024onedm} model as a text-conditioned surrogate generator. Unlike FSMG, which targets DreamBooth personalization and requires a subject-specific surrogate model for each protected subject, \Name directly optimizes against a frozen one-shot handwriting generator without additional per-writer training.

Given a set of $K$ pseudo content texts $C=\{c_1,\ldots,c_K\}$, we pre-generate and cache clean pseudo generations $\widetilde{y}_{c}=G_{\theta}(c,x)$ for each $c\in C$, where $G_{\theta}$ denotes the frozen surrogate generator. At each optimization step, we encode each cached pseudo generation $\widetilde{y}_{c}$ into the latent space, sample a diffusion timestep and noise, and evaluate the frozen One-DM denoising process under content condition $c$. In this process, the clean reference $x$ in the reference-dependent style and stroke conditions is replaced with the current protected reference $x_{\mathrm{pro}}$.

Let $\mathcal{L}_{\mathrm{den}}(\widetilde{y}_{c},c,x_{\mathrm{pro}})$ denote the resulting denoising prediction error between the predicted and sampled noise. We define the surrogate generation objective as
\begin{equation}
\label{eq:generation_objective}
J_{\mathrm{gen}}
=
\frac{1}{K}
\sum_{c\in C}
\mathcal{L}_{\mathrm{den}}
\left(
\widetilde{y}_{c},
c,
x_{\mathrm{pro}}
\right).
\end{equation}
Maximizing $J_{\mathrm{gen}}$ makes the protected reference less compatible with the clean-reference generation behavior of the target writer. Optimizing over multiple pseudo content texts further prevents the perturbation from overfitting to a particular word.

All surrogate-model parameters remain frozen, and gradients are propagated only to $\delta$. The final objective combines style displacement, surrogate generation disruption, and perceptual stealthiness:
\begin{equation}
\label{eq:sap_objective}
\max_{\delta:\|\delta\|_{\infty}\leq\epsilon}
\quad
J
=
J_{\mathrm{style}}
+
\lambda_{\mathrm{gen}}J_{\mathrm{gen}}
-
\lambda_{\mathrm{perc}}P_{\mathrm{perc}}.
\end{equation}

Starting from a masked initialization $\delta^{(0)}$, we optimize $\delta$ using PGD-style projected gradient ascent~\cite{madry2018towards}. At iteration $k$, the perturbation is updated as
\begin{equation}
\label{eq:pgd_update}
\delta^{k+1}
=
\Pi_{\epsilon}
\left(
M(x)\odot
\left[
\delta^{k}
+
\alpha\,\mathrm{sign}
\left(
\nabla_{\delta^{k}}J
\right)
\right]
\right).
\end{equation}
where $\alpha$ is the step size and $\Pi_{\epsilon}$ denotes projection onto the $\ell_{\infty}$ ball of radius $\epsilon$. The mask constrains every perturbation update to handwriting-aware regions, while the projection enforces the perturbation budget. After each update, we recompute $x_{\mathrm{pro}}$ using Eq.~\ref{eq:protected_reference}, which also clips it to the valid image range.

\begin{table*}[!t]
\centering
\small
\setlength{\tabcolsep}{3.5pt}
\renewcommand{\arraystretch}{1.08}

\caption{\textbf{Main results on One-DM.} Top1/Top5 denote target-writer retrieval rates; BgAbs and BgE denote background average absolute perturbation and background energy ratio.}
\label{Table1}
\begin{tabular*}{\textwidth}{@{\extracolsep{\fill}}lcccccccccccc}
\toprule
&
\multicolumn{2}{c}{\textbf{ResNet50}} &
\multicolumn{2}{c}{\textbf{DeepWriter}} &
\multicolumn{5}{c}{\textbf{Reference Stealth}} &
\multicolumn{1}{c}{\textbf{Content}} &
\multicolumn{2}{c}{\textbf{Generation Quality}} \\
\cmidrule(lr){2-3}
\cmidrule(lr){4-5}
\cmidrule(lr){6-10}
\cmidrule(lr){11-11}
\cmidrule(lr){12-13}
\textbf{Method} &
\textbf{Top1 $\downarrow$} &
\textbf{Top5 $\downarrow$} &
\textbf{Top1 $\downarrow$} &
\textbf{Top5 $\downarrow$} &
\textbf{LPIPS $\downarrow$} &
\textbf{PSNR $\uparrow$} &
\textbf{SSIM $\uparrow$} &
\textbf{BgAbs $\downarrow$} &
\textbf{BgE $\downarrow$} &
\textbf{CER $\downarrow$} &
\textbf{HWD $\downarrow$} &
\textbf{FID $\downarrow$} \\
\midrule
Clean &
13.20 & 38.73 &
10.68 & 34.31 &
-- & -- & -- &
-- & -- &
15.51 & 1.862 & 123.43 \\

DiffusionGuard &
4.08 & 15.82 &
4.03 & 15.47 &
0.0096 & 37.98 & 0.9746 &
0.0046 & 63.05 &
18.36 & \textbf{1.979} & 131.48 \\

PhotoGuard &
3.84 & 15.66 &
3.92 & 15.32 &
0.0087 & \textbf{38.09} & 0.9761 &
0.0043 & 59.60 &
18.20 & 1.987 & 131.46 \\

StyleGuard &
2.60 & 10.20 &
2.42 & 10.06 &
0.1249 & 20.75 & 0.9111 &
0.0170 & 55.44 &
17.49 & 2.053 & 133.78 \\

Anti-DreamBooth &
4.13 & 15.35 &
3.83 & 14.85 &
0.0138 & 36.64 & 0.9532 &
0.0122 & 69.73 &
18.61 & 1.989 & \textbf{130.95} \\
\midrule
\textbf{\Name} &
\textbf{2.12} & \textbf{9.21} &
\textbf{1.93} & \textbf{8.37} &
\textbf{0.0078} & 33.44 & \textbf{0.9856} &
\textbf{0.0000} & \textbf{0.00} &
\textbf{17.40} & 2.157 & 137.17 \\
\bottomrule
\end{tabular*}
\end{table*}

\begin{figure*}[t]
    \centering
    \includegraphics[width=\textwidth]{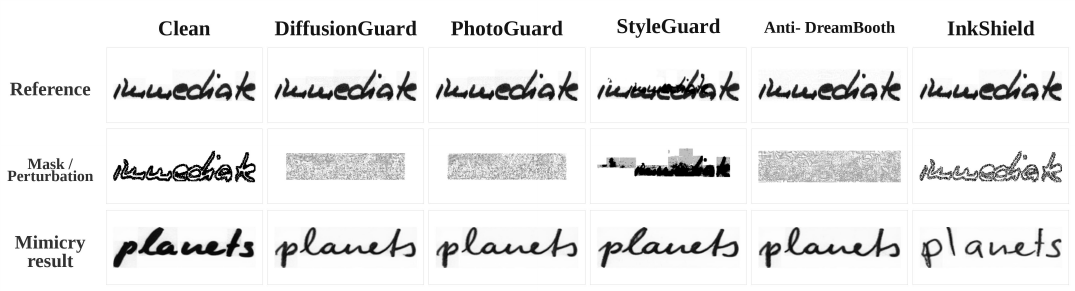}
    \caption{\textbf{Qualitative comparison of reference protection methods on
One-DM.} All methods use the same clean reference from one target writer
and generate the requested word \textit{planets}. The first column shows the shared clean reference, \Name edge mask, and clean generation; the remaining columns show the
corresponding protected references, perturbations, and mimicry results.}
    \label{fig:main_visual}
    
\end{figure*}

\section{Experimental Setup}

\noindent\textbf{Dataset.}
We evaluate \Name on the IAM handwriting dataset~\cite{marti2002iam}, which contains 62,857 English word images from 500 writers. Following prior work~\cite{bhunia2021handwriting,kang2020ganwriting,pippi2023visual}, we split the writers into 339 training writers and 161 test writers. The training split is used to train the writer evaluators introduced below, while all protection and generation experiments are conducted on the held-out test writers. To instantiate the reference-abuse threat in a controlled setting, we conduct experiments under the OOV-U protocol, in which both the test writers and content words are unseen during training. For each held-out target writer, we evaluate generation on 80 OOV-U content words, yielding $161\times80=12{,}880$ generated word images per method. For the unified evaluation protocol, word images are resized to a height of 64 pixels while preserving their aspect ratios.

\noindent\textbf{Evaluation Metrics.}
We evaluate \Name from four complementary perspectives: target-writer mimicry suppression, reference stealth, content preservation, and generation quality. These perspectives must be considered jointly: a reduction in target-writer similarity alone does not constitute successful protection if it is achieved by visibly corrupting the released reference or severely degrading the generated handwriting.

\begin{itemize}
\setlength{\itemsep}{0pt}
\setlength{\parsep}{0pt}
\setlength{\topsep}{2pt}

\item \emph{Target-writer mimicry suppression:}
We assess target-writer mimicry using two independently trained writer evaluators, \textbf{ResNet50}~\cite{he2016deep} and \textbf{DeepWriter}~\cite{xing2016deepwriter}. Using penultimate-layer embeddings, we retrieve gallery writers for each generated image and report target-writer Top-1 and Top-5 retrieval rates; lower values indicate stronger mimicry suppression. Training details are provided in Appendix.

\item \emph{Reference stealth:}
We compare each protected reference with its clean counterpart using \textbf{LPIPS}~\cite{zhang2018unreasonable}, \textbf{PSNR}~\cite{huynh2008scope}, and \textbf{SSIM}~\cite{wang2004image}. Lower LPIPS indicates smaller perceptual distortion, whereas higher PSNR and SSIM indicate better pixel-level and structural preservation, respectively. We further report background average absolute perturbation \textbf{(BgAbs)} and background energy ratio \textbf{(BgE)}. Lower BgAbs and BgE indicate that perturbations are concentrated on handwriting regions rather than blank backgrounds.

\item \emph{Content preservation:}
We use TrOCR~\cite{li2023trocr} to transcribe generated word images and compute the character error rate \textbf{(CER)} with respect to the requested content text. Lower CER indicates better textual fidelity and provides an automated proxy for generated-text legibility.

\item \emph{Generation quality:}
We report writer-level \textbf{HWD}~\cite{pippi2023hwd} and \textbf{FID}~\cite{heusel2017gans} by comparing generated images with real IAM test images from the corresponding target writer in an unpaired manner. Lower HWD and FID indicate closer distributional agreement with real handwriting. Since the protection is intended to weaken target-writer style similarity, these measures serve to detect severe generation degradation rather than as direct protection objectives.
\end{itemize}

\noindent\textbf{Generation Models.}
We use One-DM~\cite{dai2024onedm} as the surrogate model and primary evaluation generator. One-DM and CONSTANT~\cite{le2026constant} each condition generation on one fixed reference image per target writer. DiffusionPen~\cite{nikolaidou2024diffusionpen} follows its multi-shot setting and uses five references per target writer; we independently protect all five references with \Name. This protocol evaluates whether protected references transfer across generators with different reference-conditioning schemes.

\noindent\textbf{Comparison Methods.}
For the primary One-DM evaluation, we compare \Name with four representative image-protection methods: DiffusionGuard~\cite{choi2025diffusionguard}, PhotoGuard~\cite{salman2023photoguard}, StyleGuard~\cite{li2025styleguard}, and Anti-DreamBooth~\cite{vanle2023antidreambooth}. These methods respectively cover protection against diffusion-based image editing, style mimicry, and personalized image synthesis. Each baseline is applied to the same clean handwriting references and evaluated using the same One-DM generation protocol and evaluation metrics as \Name.

\noindent\textbf{Implementation Details.}
\Name performs PGD-style projected gradient ascent in normalized grayscale image space with a masked random start. We use the P95 decoy selection ($\rho=0.95$), the handwriting edge mask, a perturbation budget of $\epsilon=16/255$, a step size of $\alpha=2/255$, and $80$ optimization steps. The surrogate-generation objective uses $K=4$ cached pseudo content texts with $\lambda_{\mathrm{gen}}=3.0$. We use an LPIPS soft penalty with threshold $\tau=0.005$. The surrogate is a frozen One-DM model with a Stable Diffusion v1.5 backbone, whose parameters remain fixed during optimization. The component-wise style weights and the remaining loss weights are provided in the Appendix.

\subsection{Main Results}

Table~\ref{Table1} reports the main results on one-shot handwriting generation with One-DM. \Name achieves the strongest target-writer mimicry suppression under both independent writer evaluators. On ResNet50, it reduces target-writer Top-1/Top-5 retrieval from $13.20\%$/$38.73\%$ for clean generation to $2.12\%$/$9.21\%$. On DeepWriter, the corresponding rates decrease from $10.68\%$/$34.31\%$ to $1.93\%$/$8.37\%$. \Name obtains the lowest retrieval rates across all four evaluator--retrieval combinations, showing that protected references substantially reduce the attribution of downstream generations to the target writer.

\Name also achieves strong reference stealth. Its $\mathrm{BgAbs}=0.0000$ and $\mathrm{BgE}=0.00$ indicate that perturbations are confined to the handwriting-aware mask without affecting blank background regions. It further achieves the lowest LPIPS ($0.0078$) and the highest SSIM ($0.9856$) among all compared methods. These results show that \Name suppresses mimicry through localized, low-visibility perturbations rather than broad reference corruption.

\Name maintains generated-text readability, achieving a CER of $17.40\%$, close to the clean result of $15.51\%$ and lower than several competing defenses. Its HWD and FID increase relative to clean generation, as expected when outputs depart from the target-writer distribution, and serve only to diagnose severe generation degradation. Figure~\ref{fig:main_visual} shows that \Name concentrates perturbations near stroke boundaries while preserving the requested content word.

\subsection{Cross-model Transferability}

\begin{figure}[t]
    \centering
    \includegraphics[width=\columnwidth]{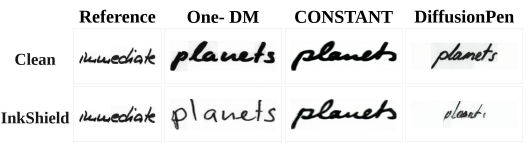}
    \caption{\textbf{Cross-model transferability.}
    Qualitative generations from One-DM, CONSTANT, and DiffusionPen for the requested word \textit{planets}. DiffusionPen uses five independently \Name-protected references; one representative reference is shown.}
    \label{fig:transfer_visual}
\end{figure}


Figure~\ref{fig:transfer_visual} compares generations obtained from clean and \Name-protected references. Clean references enable all three generators to reproduce target-writer-like characteristics, whereas protected references induce visible style departures while preserving the requested word. The effect is strongest on the One-DM surrogate, but the changes observed for CONSTANT and DiffusionPen demonstrate transferability across generators with different reference-conditioning schemes. For DiffusionPen, we independently protect all five conditioning references, indicating that the defense remains applicable when generation uses multiple style examples. Detailed quantitative cross-model results are provided in the Appendix.

\subsection{Ablation Study}

Table~\ref{tab:ablation} reports ablations of the \Name components. RN and DW denote the ResNet50 and DeepWriter evaluators, and RN@1/RN@5 and DW@1/DW@5 denote target-writer Top-1/Top-5 retrieval-rate reductions relative to clean generation. D-Rank is the average reduction in the selected decoy writer's retrieval rank across the two evaluators. ``- Dec.'', ``- Tgt.'', and ``- Gen.'' remove decoy attraction, target-writer repulsion, and surrogate-generation optimization, respectively.

Removing either decoy attraction or target-writer repulsion weakens mimicry suppression, confirming that both terms are needed for effective style displacement. Removing the surrogate-generation objective produces the largest degradation: RN@5 decreases from $29.52$ to $26.75$, and DW@5 decreases from $25.94$ to $23.46$. This result shows that static style representations alone are insufficient for the one-shot handwriting threat.

The global-mask variant yields stronger suppression (RN@5: $32.83$; DW@5: $28.92$), but incurs a substantial stealth cost: LPIPS increases from $0.0078$ to $0.0915$, and BgE rises from $0.00$ to $67.90$. It therefore serves as an aggressive reference point rather than a practical defense. Overall, \Name achieves a more favorable protection--stealth trade-off through decoy-guided displacement, surrogate-generation optimization, and edge-constrained perturbations.

\begin{table}[t]
\centering
\scriptsize
\setlength{\tabcolsep}{2.0pt}
\renewcommand{\arraystretch}{1.06}
\caption{\textbf{Ablation study of \Name on One-DM.} RN and DW denote
ResNet50 and DeepWriter, respectively. $\Delta$Top1/$\Delta$Top5 denote
target-writer retrieval-rate reductions relative to clean generation.
D-Rank denotes the selected decoy writer's rank decrease.}
\label{tab:ablation}
\begin{tabular*}{\columnwidth}{@{\extracolsep{\fill}}lccccccc}
\toprule
\textbf{Variant} &
\textbf{RN@1 $\uparrow$} &
\textbf{RN@5 $\uparrow$} &
\textbf{DW@1 $\uparrow$} &
\textbf{DW@5 $\uparrow$} &
\textbf{D-Rank $\uparrow$} &
\textbf{LPIPS $\downarrow$} &
\textbf{BgE $\downarrow$} \\
\midrule
Global mask &
11.77 & 32.83 &
9.70 & 28.92 &
21.51 & 0.0915 & 67.90 \\
\midrule
\textbf{\Name} &
\textbf{11.08} & \textbf{29.52} &
\textbf{8.75} & \textbf{25.94} &
\textbf{16.43} & \textbf{0.0078} & \textbf{0.00} \\
- Dec. &
10.88 & 28.25 &
8.35 & 24.27 &
10.79 & 0.0136 & 0.00 \\
- Tgt. &
10.58 & 28.67 &
8.43 & 24.98 &
15.27 & 0.0140 & 0.00 \\
- Gen. &
9.98 & 26.75 &
8.18 & 23.46 &
11.99 & 0.0150 & 0.00 \\
\bottomrule
\end{tabular*}
\end{table}

\subsection{Hyperparameter Sensitivity}

\begin{figure}[t]
    \centering
    \includegraphics[width=\columnwidth]{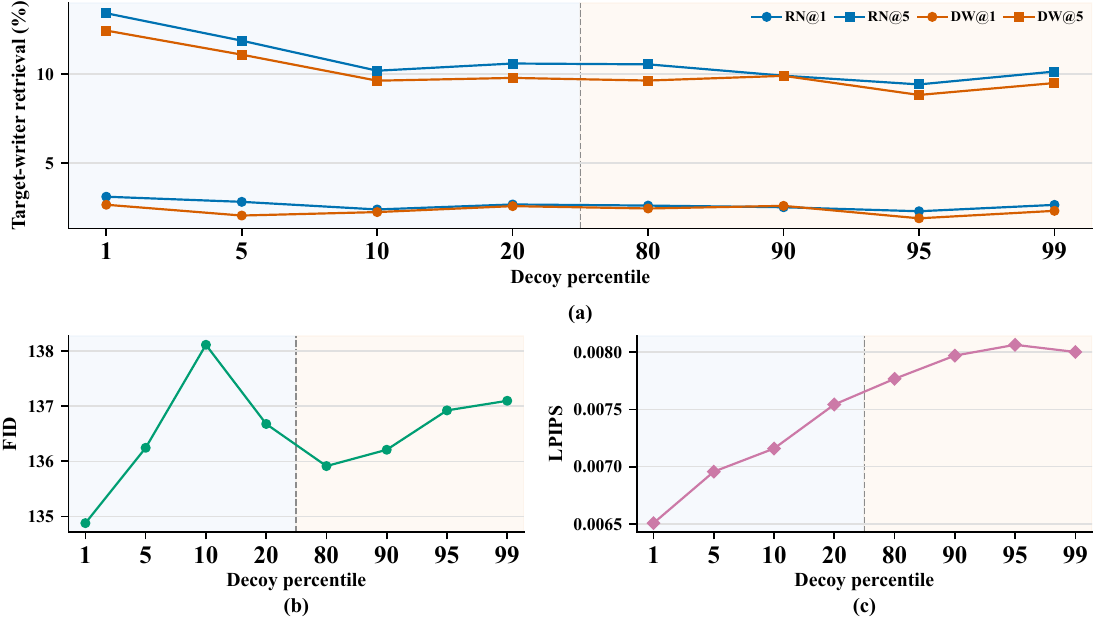}
    \caption{\textbf{Sensitivity to the decoy percentile.}
    Target-writer retrieval rates, FID, and LPIPS across decoy percentiles are shown in (a), (b), and (c), respectively. P95 provides the selected protection--stealth trade-off.}
    \label{fig:target_sensitivity}
\end{figure}

Figure~\ref{fig:target_sensitivity} evaluates the decoy percentile while keeping all other settings fixed. Increasing the percentile generally lowers target-writer Top-1 and Top-5 retrieval rates under both ResNet50 and DeepWriter, since more distant decoys provide stronger style-displacement directions. P95 achieves the lowest or near-lowest retrieval rates without relying on the single farthest decoy. More extreme decoys provide limited additional suppression but increase LPIPS and lead to a less favorable FID trend. We therefore use P95 as the default, as it provides strong mimicry suppression with localized, low-visibility perturbations. Additional sensitivity results for the perturbation budget, optimization steps, number of pseudo content texts, and perceptual-constraint parameters are provided in the Appendix.

\section{Conclusion}


In this paper, we study proactive reference-side protection against unauthorized handwriting style mimicry. We propose \Name, which protects handwriting references before release through decoy-guided style displacement, stroke-edge constraints, and frozen surrogate optimization. Experiments on IAM show that \Name suppresses target-writer mimicry while preserving reference readability and visual utility, with transfer to CONSTANT and DiffusionPen. These results demonstrate a practical defense against unauthorized style reuse of publicly shared handwriting.



\bibliography{main}

@inproceedings{dai2024onedm,
author    = {Dai, Gang and Zhang, Yifan and Ke, Quhui and Guo, Qiangya and Huang, Shuangping},
title     = {{One-DM}: One-Shot Diffusion Mimicker for Handwritten Text Generation},
booktitle = {Proceedings of the European Conference on Computer Vision},
year      = {2024}
}

@inproceedings{le2026constant,
author    = {Le, Anh-Duy and Pham, Van-Linh and Vo, Thanh-Nam and Mai, Xuan Toan and Tran, Tuan-Anh},
title     = {{CONSTANT}: Towards High-Quality One-Shot Handwriting Generation with Patch Contrastive Enhancement and Style-Aware Quantization},
booktitle = {Proceedings of the IEEE/CVF Winter Conference on Applications of Computer Vision},
year      = {2026}
}

@inproceedings{nikolaidou2024diffusionpen,
author    = {Nikolaidou, Konstantina and Retsinas, George and Sfikas, Giorgos and Liwicki, Marcus},
title     = {{DiffusionPen}: Towards Controlling the Style of Handwritten Text Generation},
booktitle = {Proceedings of the European Conference on Computer Vision Workshops},
year      = {2024}
}

@inproceedings{shan2023glaze,
author    = {Shan, Shawn and Cryan, Jenna and Wenger, Emily and Zheng, Haitao and Hanocka, Rana and Zhao, Ben Y.},
title     = {Glaze: Protecting Artists from Style Mimicry by Text-to-Image Models},
booktitle = {Proceedings of the USENIX Security Symposium},
year      = {2023}
}

@misc{liang2023mist,
author        = {Liang, Chumeng and Wu, Xiaoyu},
title         = {Mist: Towards Improved Adversarial Examples for Diffusion Models},
year          = {2023},
eprint        = {2305.12683},
archivePrefix = {arXiv},
primaryClass  = {cs.CR}
}

@inproceedings{salman2023photoguard,
author    = {Salman, Hadi and Khaddaj, Alaa and Leclerc, Guillaume and Ilyas, Andrew and Madry, Aleksander},
title     = {Raising the Cost of Malicious {AI}-Powered Image Editing},
booktitle = {Proceedings of the International Conference on Machine Learning},
pages     = {29894--29918},
year      = {2023}
}

@inproceedings{choi2025diffusionguard,
author    = {Choi, June Suk and Lee, Kyungmin and Jeong, Jongheon and Xie, Saining and Shin, Jinwoo and Lee, Kimin},
title     = {DiffusionGuard: A Robust Defense Against Malicious Diffusion-Based Image Editing},
booktitle = {Proceedings of the International Conference on Learning Representations},
year      = {2025}
}

@inproceedings{vanle2023antidreambooth,
author    = {Van Le, Thanh and Phung, Hao and Nguyen, Thuan Hoang and Dao, Quan and Tran, Ngoc and Tran, Anh},
title     = {Anti-DreamBooth: Protecting Users from Personalized Text-to-Image Synthesis},
booktitle = {Proceedings of the IEEE/CVF International Conference on Computer Vision},
pages     = {2116--2127},
year      = {2023}
}

@inproceedings{liu2024metacloak,
author    = {Liu, Yixin and Fan, Chenrui and Dai, Yutong and Chen, Xun and Zhou, Pan and Sun, Lichao},
title     = {MetaCloak: Preventing Unauthorized Subject-Driven Text-to-Image Diffusion-Based Synthesis via Meta-Learning},
booktitle = {Proceedings of the IEEE/CVF Conference on Computer Vision and Pattern Recognition},
pages     = {24219--24228},
year      = {2024}
}

@inproceedings{li2025styleguard,
author    = {Li, Yanjie and Zhang, Wenxuan and Lyu, Xinqi and Liu, Yihao and Xiao, Bin},
title     = {StyleGuard: Preventing Text-to-Image-Model-Based Style Mimicry Attacks by Style Perturbations},
booktitle = {Advances in Neural Information Processing Systems},
year      = {2025}
}

@inproceedings{bhunia2021handwriting,
  author    = {Bhunia, Ankan Kumar and Khan, Salman and Cholakkal, Hisham and Anwer, Rao Muhammad and Khan, Fahad Shahbaz and Shah, Mubarak},
  title     = {Handwriting Transformers},
  booktitle = {Proceedings of the IEEE/CVF International Conference on Computer Vision},
  pages     = {1086--1094},
  year      = {2021}
}

@inproceedings{kang2020ganwriting,
  author    = {Kang, Lei and Riba, Pau and Wang, Yaxing and Rusi{\~n}ol, Mar{\c{c}}al and Forn{\'e}s, Alicia and Villegas, Mauricio},
  title     = {{GANwriting}: Content-Conditioned Generation of Styled Handwritten Word Images},
  booktitle = {Proceedings of the European Conference on Computer Vision},
  pages     = {273--289},
  year      = {2020}
}

@inproceedings{pippi2023visual,
  author    = {Pippi, Vittorio and Cascianelli, Silvia and Cucchiara, Rita},
  title     = {Handwritten Text Generation from Visual Archetypes},
  booktitle = {Proceedings of the IEEE/CVF Conference on Computer Vision and Pattern Recognition},
  pages     = {22458--22467},
  year      = {2023}
}

@article{marti2002iam,
  title={The IAM-database: an English sentence database for offline handwriting recognition},
  author={Marti, U-V and Bunke, Horst},
  journal={International journal on document analysis and recognition},
  volume={5},
  number={1},
  pages={39--46},
  year={2002},
  publisher={Springer}
}

@inproceedings{zhang2018unreasonable,
  author    = {Zhang, Richard and Isola, Phillip and Efros, Alexei A. and Shechtman, Eli and Wang, Oliver},
  title     = {The Unreasonable Effectiveness of Deep Features as a Perceptual Metric},
  booktitle = {Proceedings of the IEEE Conference on Computer Vision and Pattern Recognition},
  pages     = {586--595},
  year      = {2018}
}

@article{wang2004image,
  author  = {Wang, Zhou and Bovik, Alan C. and Sheikh, Hamid R. and Simoncelli, Eero P.},
  title   = {Image Quality Assessment: From Error Visibility to Structural Similarity},
  journal = {IEEE Transactions on Image Processing},
  volume  = {13},
  number  = {4},
  pages   = {600--612},
  year    = {2004}
}

@article{huynh2008scope,
  title={Scope of validity of PSNR in image/video quality assessment},
  author={Huynh-Thu, Quan and Ghanbari, Mohammed},
  journal={Electronics letters},
  volume={44},
  number={13},
  pages={800--801},
  year={2008},
  publisher={IET}
}

@inproceedings{li2023trocr,
  title={Trocr: Transformer-based optical character recognition with pre-trained models},
  author={Li, Minghao and Lv, Tengchao and Chen, Jingye and Cui, Lei and Lu, Yijuan and Florencio, Dinei and Zhang, Cha and Li, Zhoujun and Wei, Furu},
  booktitle={Proceedings of the AAAI conference on artificial intelligence},
  volume={37},
  number={11},
  pages={13094--13102},
  year={2023}
}

@article{pippi2023hwd,
  title={HWD: A novel evaluation score for styled handwritten text generation},
  author={Pippi, Vittorio and Quattrini, Fabio and Cascianelli, Silvia and Cucchiara, Rita},
  journal={arXiv preprint arXiv:2310.20316},
  year={2023}
}

@article{heusel2017gans,
  title={Gans trained by a two time-scale update rule converge to a local nash equilibrium},
  author={Heusel, Martin and Ramsauer, Hubert and Unterthiner, Thomas and Nessler, Bernhard and Hochreiter, Sepp},
  journal={Advances in neural information processing systems},
  volume={30},
  year={2017}
}

@article{vanherle2024vatrpp,
  title={Vatr++: Choose your words wisely for handwritten text generation},
  author={Vanherle, Bram and Pippi, Vittorio and Cascianelli, Silvia and Michiels, Nick and Van Reeth, Frank and Cucchiara, Rita},
  journal={IEEE Transactions on Pattern Analysis and Machine Intelligence},
  volume={47},
  number={2},
  pages={934--948},
  year={2024},
  publisher={IEEE}
}

@inproceedings{gan2021higan,
  author    = {Gan, Ji and Wang, Weiqiang},
  title     = {{HiGAN}: Handwriting Imitation Conditioned on Arbitrary-Length Texts and Disentangled Styles},
  booktitle = {Proceedings of the AAAI Conference on Artificial Intelligence},
  volume    = {35},
  number    = {9},
  pages     = {7484--7492},
  year      = {2021},
  doi       = {10.1609/aaai.v35i9.16917}
}

@inproceedings{nikolaidou2023wordstylist,
  author    = {Nikolaidou, Konstantina and Retsinas, George and Christlein, Vincent and Seuret, Mathias and Sfikas, Giorgos and Barney Smith, Elisa and Mokayed, Hamam and Liwicki, Marcus},
  title     = {{WordStylist}: Styled Verbatim Handwritten Text Generation with Latent Diffusion Models},
  booktitle = {Document Analysis and Recognition -- ICDAR 2023},
  series    = {Lecture Notes in Computer Science},
  volume    = {14188},
  pages     = {384--401},
  publisher = {Springer},
  year      = {2023},
  doi       = {10.1007/978-3-031-41679-8_22}
}

@inproceedings{chen2024editshield,
  author    = {Chen, Ruoxi and Jin, Haibo and Liu, Yixin and Chen, Jinyin and Wang, Haohan and Sun, Lichao},
  title     = {{EditShield}: Protecting Unauthorized Image Editing by Instruction-Guided Diffusion Models},
  booktitle = {Computer Vision -- ECCV 2024},
  series    = {Lecture Notes in Computer Science},
  volume    = {15121},
  pages     = {126--142},
  publisher = {Springer},
  year      = {2024}
}

@inproceedings{wan2024pap,
  author    = {Wan, Cong and He, Yuhang and Song, Xiang and Gong, Yihong},
  title     = {Prompt-Agnostic Adversarial Perturbation for Customized Diffusion Models},
  booktitle = {Advances in Neural Information Processing Systems},
  volume    = {37},
  year      = {2024}
}

@inproceedings{dai2025diffbrush,
  author    = {Dai, Gang and Zhang, Yifan and Qin, Yutao and Guo, Qiangya and Huang, Shuangping and Yan, Shuicheng},
  title     = {Beyond Isolated Words: Diffusion Brush for Handwritten Text-Line Generation},
  booktitle = {Proceedings of the IEEE/CVF International Conference on Computer Vision},
  pages     = {19054--19064},
  year      = {2025}
}

@inproceedings{madry2018towards,
  title={Towards Deep Learning Models Resistant to Adversarial Attacks},
  author={Madry, Aleksander and Makelov, Aleksandar and Schmidt, Ludwig and Tsipras, Dimitris and Vladu, Adrian},
  booktitle={International Conference on Learning Representations},
  year={2018}
}

@misc{sobel1968isotropic,
  author       = {Irwin Sobel and Gary Feldman},
  title        = {A 3x3 Isotropic Gradient Operator for Image Processing},
  year         = {1968},
  howpublished = {Presented at the Stanford Artificial Intelligence Project}
}

@inproceedings{he2016deep,
  title={Deep Residual Learning for Image Recognition},
  author={He, Kaiming and Zhang, Xiangyu and Ren, Shaoqing and Sun, Jian},
  booktitle={Proceedings of the IEEE Conference on Computer Vision and Pattern Recognition},
  pages={770--778},
  year={2016}
}

@article{xing2016deepwriter,
  title={DeepWriter: A Multi-Stream Deep CNN for Text-Independent Writer Identification},
  author={Xing, Linjie and Qiao, Yu},
  journal={arXiv preprint arXiv:1606.06472},
  year={2016}
}

\clearpage
\appendix

\section{A Writer Evaluator Training and Validation}
\label{app:evaluator_training}

We use two independently trained writer evaluators, \textbf{ResNet50} and \textbf{DeepWriter}, to assess target-writer mimicry. Both models are trained as writer classifiers, while evaluation uses their penultimate-layer embeddings for writer-level retrieval rather than their softmax outputs. This design measures whether a generated image remains close to the target writer in the learned writer-style space.
For both evaluators, we use the $161$ IAM test writers as the target-writer set. For each writer, five images are reserved to construct the retrieval gallery, resulting in $805$ gallery images in total. These gallery images are excluded from training and validation. The remaining real test-writer images are split independently by writer into approximately $70\%$/$10\%$/$20\%$ training, development, and held-out test subsets. We select the checkpoint with the best development Top-1 classification accuracy.

\textbf{ResNet50.}
We initialize ResNet50 from ImageNet weights. Each grayscale word image is inverted, padded to a square canvas, and resized to $224\times224$. The model is first trained for writer classification on the IAM training writers. It is then transferred to the $161$ test writers by retaining the backbone, reinitializing the $161$-way classification head, and fine-tuning on the corresponding training subset.

\textbf{DeepWriter.}
DeepWriter uses a two-stream architecture tailored to handwritten word images. Inputs preserve their aspect ratios after resizing to height $113$, and two width-$113$ strips are sampled as the two streams. We first train the model for writer classification on the IAM training writers, then transfer its branches and backbone to the $161$ test writers, reinitialize the $161$-way classification head, and fine-tune using the same split protocol.

For validation, we extract penultimate-layer embeddings for held-out real images and retrieve the nearest writer among the $161$ gallery centroids. Table~\ref{tab:evaluator_clean} reports this clean held-out retrieval performance over $4{,}927$ query images. These results establish that both evaluators can reliably recognize writer identity on real IAM images before being used to assess target-writer mimicry in generated samples.

\begin{table}[!h]
\centering
\small
\setlength{\tabcolsep}{3.5pt}
\renewcommand{\arraystretch}{1.08}
\caption{\textbf{Clean held-out retrieval performance of writer evaluators on IAM test writers.} Five images per writer form the gallery ($805$ images total), and results are computed on $4{,}927$ held-out real query images. Top-1/Top-5 are embedding-retrieval rates, not softmax classification accuracies. Source similarity denotes the cosine similarity to the centroid of the true writer.}
\label{tab:evaluator_clean}
\begin{tabular}{lcccc}
\toprule
\textbf{Evaluator} &
\textbf{Top-1 $\uparrow$} &
\textbf{Top-5 $\uparrow$} &
\textbf{Median Rank $\downarrow$} &
\textbf{Source Sim. $\uparrow$} \\
\midrule
ResNet50 &
79.84 &
91.66 &
2.25 &
0.8088 \\
DeepWriter &
65.15 &
85.40 &
3.72 &
0.7921 \\
\bottomrule
\end{tabular}
\end{table}

\section{B Additional Cross-Model Transferability Results}
\label{app:cross_model_transfer}
We further evaluate whether \Name-protected references transfer to handwriting generators with different reference-conditioning mechanisms. One-DM is the frozen surrogate used during optimization, whereas CONSTANT performs one-shot generation with a distinct architecture. DiffusionPen follows its multi-shot setting, in which all five reference images of each target writer are independently protected by \Name.

Raw target-writer retrieval rates are not directly comparable across generators because their clean-reference mimicry abilities differ. We therefore report both the clean and protected retrieval rates, together with chance-adjusted suppression. Let $R^{\mathrm{C}}_{G,E}@K$ and $R^{\mathrm{P}}_{G,E}@K$ denote the target-writer Top-$K$ retrieval rates for generator $G$ under evaluator $E$ using clean and protected references, respectively. With $N=161$ candidate writers, the chance retrieval rate is $K/N$. We define
\begin{equation}
\label{eq:chance_adjusted_suppression}
\mathrm{Supp}_{G,E}@K
=
\frac{
R^{\mathrm{C}}_{G,E}@K
-
R^{\mathrm{P}}_{G,E}@K
}{
R^{\mathrm{C}}_{G,E}@K
-
K/N
}
\times 100\%.
\end{equation}
This metric measures the fraction of target-writer retrievability above the random baseline that is removed by protection.

Table~\ref{tab:cross_model_transfer} shows that \Name provides the strongest suppression on the One-DM surrogate, removing $82.86\%$--$88.08\%$ of above-chance target-writer retrieval across the two evaluators and retrieval levels. On CONSTANT and DiffusionPen, \Name still removes approximately $28\%$--$39\%$ of above-chance retrieval. These results indicate cross-model transferability: although the defense effect is weaker than that on the surrogate, protected references consistently reduce target-writer mimicry across one-shot and multi-shot generators.

\begin{table}[!h]
\centering
\small
\setlength{\tabcolsep}{3.2pt}
\renewcommand{\arraystretch}{1.08}
\caption{\textbf{Additional cross-model transferability results.}
C and P denote clean-reference and \Name-protected-reference generation,
respectively. Supp. denotes chance-adjusted target-writer suppression
defined in Eq.~\ref{eq:chance_adjusted_suppression}. All entries are
percentages.}
\label{tab:cross_model_transfer}
\begin{tabular}{llccc}
\toprule
\multicolumn{5}{c}{\textbf{(a) Top-1 Retrieval}} \\
\cmidrule(lr){1-5}
\textbf{Generator} &
\textbf{Evaluator} &
\textbf{C $\downarrow$} &
\textbf{P $\downarrow$} &
\textbf{Supp. $\uparrow$} \\
\midrule
One-DM       & ResNet50    & 13.20 & 2.12 & 88.08 \\
One-DM       & DeepWriter  & 10.68 & 1.93 & 86.99 \\
\midrule
CONSTANT     & ResNet50    & 6.34  & 4.12 & 38.82 \\
CONSTANT     & DeepWriter  & 5.95  & 4.29 & 31.15 \\
\midrule
DiffusionPen & ResNet50    & 6.27  & 4.35 & 33.99 \\
DiffusionPen & DeepWriter  & 5.74  & 3.87 & 36.53 \\
\midrule
\multicolumn{5}{c}{\textbf{(b) Top-5 Retrieval}} \\
\cmidrule(lr){1-5}
\textbf{Generator} &
\textbf{Evaluator} &
\textbf{C $\downarrow$} &
\textbf{P $\downarrow$} &
\textbf{Supp. $\uparrow$} \\
\midrule
One-DM       & ResNet50    & 38.73 & 9.21  & 82.86 \\
One-DM       & DeepWriter  & 34.31 & 8.37  & 83.13 \\
\midrule
CONSTANT     & ResNet50    & 24.72 & 18.77 & 27.53 \\
CONSTANT     & DeepWriter  & 21.87 & 16.05 & 31.02 \\
\midrule
DiffusionPen & ResNet50    & 20.09 & 14.98 & 30.09 \\
DiffusionPen & DeepWriter  & 18.14 & 13.39 & 31.59 \\
\bottomrule
\end{tabular}
\end{table}

\section{C Supplementary Experimental Results}
\label{app:hyperparameter_sensitivity}
We provide additional sensitivity analyses for the hyperparameters used by
\Name. These experiments complement the decoy-percentile study reported in the
main paper. In each sweep, we vary only one hyperparameter and keep the
remaining settings fixed to the default configuration. We evaluate the
resulting trade-off among target-writer mimicry suppression, reference
stealth, generation quality, and optimization cost. Unless otherwise stated,
the default configuration uses $\epsilon=16/255$, $80$ optimization steps,
$K=4$ cached pseudo content texts, LPIPS threshold $\tau=0.005$, and LPIPS
penalty weight $\lambda_{\mathrm{perc}}=1$.

\subsection{C.1 Perturbation Budget}

\begin{figure}[!h]
    \centering
    \includegraphics[width=\columnwidth]{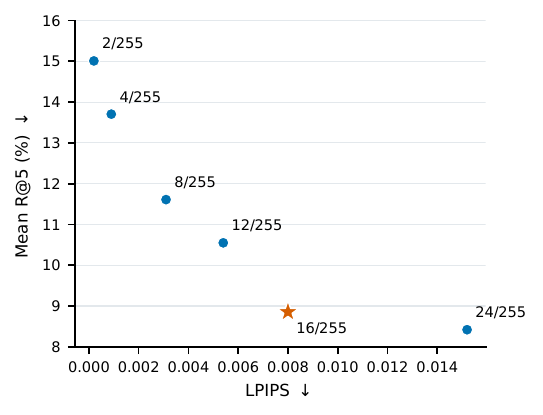}
    \caption{\textbf{Sensitivity to the perturbation budget $\epsilon$.}
    Each point is annotated with its budget. The horizontal axis denotes
    LPIPS and the vertical axis denotes mean Top-5 target-writer retrieval;
    both are lower-is-better. The orange star marks the selected default
    $\epsilon=16/255$.}
    \label{fig:epsilon_sensitivity}
\end{figure}

Figure~\ref{fig:epsilon_sensitivity} analyzes how the perturbation budget
$\epsilon$ controls the available capacity for defense-side style
displacement. Small budgets preserve a reference that is extremely close to
the clean input, but they constrain \Name's ability to modify the
style-sensitive stroke-boundary regions used by downstream generators.
Consequently, target-writer retrieval remains relatively high at the
smallest tested budgets.

Increasing $\epsilon$ generally lowers mean Top-5 target-writer retrieval,
showing that a larger feasible perturbation region enables stronger mimicry
suppression. This benefit is accompanied by higher LPIPS, which indicates a
larger perceptual deviation from the clean reference. The sweep therefore
reveals a clear suppression--stealth trade-off rather than a monotonic
preference for the largest possible budget.

We select $\epsilon=16/255$ because it lies near the favorable knee of this
trade-off. Compared with smaller budgets, it substantially reduces
target-writer retrieval; compared with $\epsilon=24/255$, it retains a lower
LPIPS while providing similar protection. This choice is therefore used in
all main experiments as a practical low-visibility operating point.

\subsection{C.2 Number of Optimization Steps}

\begin{figure}[!h]
    \centering
    \includegraphics[width=\columnwidth]{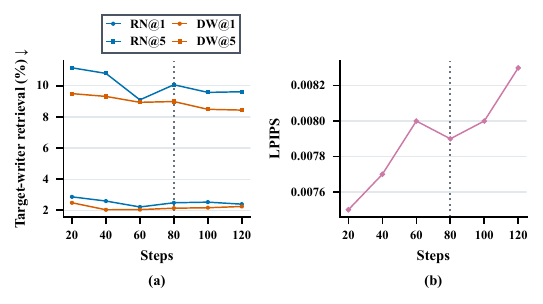}
    \caption{\textbf{Sensitivity to the number of optimization steps.}
    (a) Target-writer retrieval rates for ResNet50 (RN) and DeepWriter (DW).
    (b) LPIPS. The vertical dashed line indicates the selected default of
    $80$ steps.}
    \label{fig:step_sensitivity}
\end{figure}

Figure~\ref{fig:step_sensitivity} studies the number of projected-gradient
optimization steps. At low step counts, the perturbation has insufficient
opportunity to jointly optimize the decoy-guided style objective and the
generation-aware surrogate objective. As the step count increases,
target-writer retrieval decreases for both ResNet50 and DeepWriter, showing
that additional optimization improves suppression during the early stage.

The retrieval curves become substantially flatter around $80$ steps. Beyond
this point, additional iterations yield only limited and inconsistent
changes across the two writer evaluators. In contrast, LPIPS exhibits a
gradual upward trend as optimization proceeds, indicating that longer runs
consume more of the perceptual budget without consistently improving
target-writer mimicry suppression.

We therefore choose $80$ steps as the default. This setting reaches the
stable region of the retrieval curves while avoiding the extra runtime and
perceptual deviation associated with longer optimization. The result also
shows that \Name does not rely on an excessively long attack schedule to
obtain its main protection effect.

\begin{figure*}[!h]
\centering

\begin{minipage}[t]{0.48\textwidth}
\centering
\vspace{0pt}
\includegraphics[width=\linewidth]{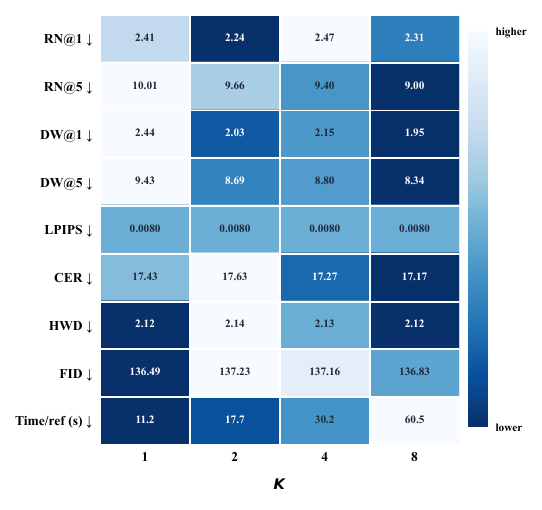}
\captionof{figure}{\textbf{Sensitivity to the number of pseudo content texts $K$.}
Cell annotations show measured values. Colors are normalized within each
metric row, with darker cells indicating lower values. The final row reports
the measured optimization time per reference sample.}
\label{fig:k_sensitivity}
\end{minipage}
\hfill
\begin{minipage}[t]{0.48\textwidth}
\centering
\vspace{0pt}
\includegraphics[width=\linewidth]{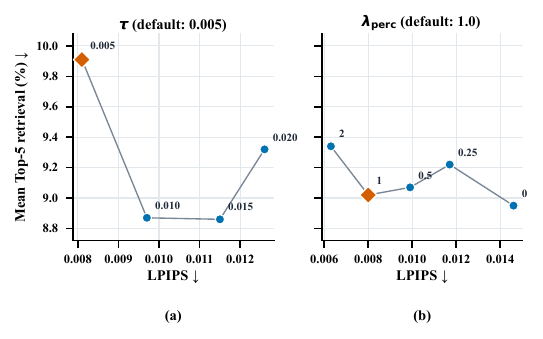}
\captionof{figure}{\textbf{Sensitivity to perceptual-constraint parameters.} (a) Effect of the LPIPS threshold $\tau$; smaller $\tau$ activates the hinge-style perceptual penalty at smaller perceptual deviations. (b) Effect of the LPIPS penalty weight $\lambda_{\mathrm{perc}}$; setting $\lambda_{\mathrm{perc}}=0$ removes the perceptual regularization. The horizontal axis is LPIPS and the vertical axis is mean Top-5 target-writer retrieval; both are lower-is-better. Orange diamonds mark the selected default settings.}
\label{fig:perceptual_sensitivity}
\end{minipage}

\end{figure*}

\subsection{C.3 Number of Pseudo Content Texts}

Figure~\ref{fig:k_sensitivity} evaluates the number $K$ of cached clean
pseudo content texts used by the generation-aware surrogate objective.
Using multiple pseudo texts exposes the protected reference to diverse
content conditions during optimization, reducing the risk that \Name
over-specializes to a single generated word or text instance.

Using multiple pseudo texts generally improves mimicry suppression relative
to a single pseudo text while preserving broadly similar LPIPS, CER, HWD,
and FID values. This pattern indicates that diverse content conditions
strengthen the generality of the surrogate objective without substantially
degrading reference stealth or generated-content fidelity.

The marginal benefit becomes small beyond $K=4$. In contrast, the measured
optimization time per reference increases markedly as more pseudo texts are
included, because each text requires an additional surrogate-loss
evaluation. We therefore use $K=4$ as the default efficiency--effectiveness
trade-off: it captures useful content diversity while avoiding the
substantial runtime cost of $K=8$.

\subsection{C.4 Perceptual-Constraint Parameters}

Figure~\ref{fig:perceptual_sensitivity}(a) analyzes the LPIPS threshold
$\tau$, which determines when the hinge-style perceptual penalty becomes
active. With
$P_{\mathrm{perc}}=\max(0,\operatorname{LPIPS}(x_{\mathrm{pro}},x)-\tau)$,
setting $\tau=0$ makes the penalty active for any nonzero perceptual
deviation. Thus, smaller thresholds impose a stricter preference for
reference-level visual similarity, whereas larger thresholds allow more
perceptual deviation before the penalty is activated.

The selected threshold $\tau=0.005$ provides a balanced operating point.
It allows \Name to modify localized handwriting regions sufficiently to
reduce target-writer retrieval, while discouraging unnecessary perceptual
deviation. These results support a thresholded soft perceptual penalty
rather than a strict hard constraint throughout optimization.

Figure~\ref{fig:perceptual_sensitivity}(b) varies the penalty weight
$\lambda_{\mathrm{perc}}$, which determines the relative importance of the
LPIPS term after the threshold is exceeded. Setting
$\lambda_{\mathrm{perc}}=0$ removes this stealth-oriented regularization,
whereas an overly large value causes the optimizer to prioritize visual
similarity over target-writer mimicry suppression. The default
$\lambda_{\mathrm{perc}}=1$ provides the most balanced observed trade-off,
maintaining low LPIPS while preserving strong suppression of target-writer
retrieval.

\end{document}